\documentclass[a4paper]{ESASPCS13Style}
\usepackage{epsfig}

\begin{document}

\title{High Resolution X-ray Spectroscopy of Pre-Main-Sequence Stars: \object{TWA~5} and \object{PZ~Tel}}

\author{C.~Argiroffi\inst{1} \and J.~J.~Drake\inst{2} \and F.~R.~Harnden\inst{2} \and A.~Maggio\inst{3} \and G.~Peres\inst{1} \and S.~Sciortino\inst{3} \and B.~Stelzer\inst{3} \and R.~Neuh\"auser\inst{4}}
\institute{Dipartimento di Scienze Fisiche ed Astronomiche, Sezione di Astronomia, Universit\`a di Palermo, Piazza del Parlamento 1, 90134 Palermo, Italy \and Smithsonian Astrophysical Observatory, 60 Garden Street, Cambridge, MA 02138 \and INAF - Osservatorio Astronomico di Palermo, Piazza del Parlamento 1, 90134 Palermo, Italy \and Astrophysikalisches Institut und Universit\"ats-Sternwarte, Schillerg\"asschen 2-3, D-07745 Jena, Germany}

\maketitle 

\begin{abstract}

We report on the analysis of high resolution X-ray spectra of two pre-main-sequence stars: \object{TWA~5} (observed with {\em XMM-Newton}) and \object{PZ~Telescopii} (observed with Chandra/HETGS). \object{TWA~5} is a classical T~Tauri star in the TW Hydrae association while \object{PZ~Tel} is a rapidly rotating weak-lined T~Tauri star in the $\beta$-Pictoris moving group. For both stars we have reconstructed the emission measure distribution and derived the coronal abundances to check for possible patterns of the abundances related to the first ionization potential of the various elements. We have also derived estimates of the plasma density from the analysis of the He-like triplets. We compare the characteristics of our targets with those of other pre-main sequence stars previously analyzed by other authors: \object{TW~Hya}, \object{HD~98800} and \object{HD~283572}. Our findings suggest that X-ray emission from classical T~Tauri and weak-lined T~Tauri stars is produced in all cases by magnetically-heated coronae, except for \object{TW~Hya} which has unique plasma temperatures and densities. Moreover we derive that \object{TWA~5} has the same peculiar Ne/Fe abundance ratio as \object{TW~Hya}.

\end{abstract}

\section{Introduction}
\label{intro}

The aim of the present work is to investigate the evolution of stellar X-ray emission during pre-main-sequence (PMS) stages. To address this issue we have analyzed the high resolution X-ray spectra of two PMS stars: the classical T~Tauri star (CTTS) \object{TWA~5}, and the weak-lined T~Tauri star (WTTS) \object{PZ~Tel} (\cite{ArgiroffiDrake2004}). These two stars allow us to probe different evolutionary phases. Mass accretion is still present in \object{TWA~5} as shown by its asymmetric and broad H$\alpha$ emission (\cite{MohantyJayawardhana2003}). \cite*{JayawardhanaHartmann1999} suggested a possible infrared excess but this finding has been ruled out by subsequent studies (\cite{MetchevHillenbrand2004,WeinbergeBecklinr2004}), indicating that \object{TWA~5} no longer harbors large amount of dust. The accretion process has just ended in \object{PZ~Tel} which has a filled in H$\alpha$ line (\cite{SoderblomKing1998}). The analysis of the X-ray emission allows us to infer the characteristics of the emitting plasma including the thermal structure, the electron densities, and the elemental abundances.

\begin{table}[!b]
\caption{Characteristics of the star sample. Negative values of H$\alpha$ equivalent width indicate an emission line.}
\label{tab1}
\begin{center}
\leavevmode
\scriptsize
\begin{tabular}{lcccrr}
\hline
Name      & Mass         & Spectral & Age      & \multicolumn{1}{c}{EW(H$\alpha$)} & \multicolumn{1}{c}{Distance} \\
          & ($M_{\odot}$)& Type     & (Myr)    & \multicolumn{1}{c}{(\AA)}         & \multicolumn{1}{c}{(pc)} \\
\hline
\object{TW~Hya}    & $\sim0.7$   & K7       & $\sim10$ & -220.00 \hspace{0.01\textwidth} & 56 \hspace{0.015\textwidth} \\
\object{TWA~5}     & $\sim0.5$   & M1.5     & $\sim10$ &  -13.64 \hspace{0.01\textwidth} & 55 \hspace{0.015\textwidth} \\
\object{HD~98800}  & $\sim1$     & K5+K7    & $\sim10$ &    0.00 \hspace{0.01\textwidth} & 47 \hspace{0.015\textwidth} \\
\object{PZ~Tel}    & $\sim1$     & K0       & $\sim12$ &    0.63 \hspace{0.01\textwidth} & 50 \hspace{0.015\textwidth} \\
\object{HD~283572} & $\sim1.6$   & G2       & $\sim10$ &    1.12 \hspace{0.01\textwidth} & 128 \hspace{0.015\textwidth} \\
\hline
\end{tabular}
\end{center}
\end{table}

High resolution X-ray spectroscopy studies have been performed for other PMS stars, and in particular for the CTTS \object{TW~Hya} (\cite{KastnerHuenemoerder2002,StelzerSchmitt2004}), and for the two WTTSs \object{HD~98800} (\cite{KastnerHuenemoerder2004}) and \object{HD~283572} (\cite{ScelsiMaggio2004a}). The analysis of \object{TWA~5} and \object{PZ~Tel} fills the gap between stars characterized by strong H$\alpha$ emission lines, like \object{TW~Hya}, and active stars near the zero age main sequence (ZAMS) with H$\alpha$ in absorption.

\begin{figure*}[!t]
\begin{center}
\resizebox{\hsize}{!}{\includegraphics{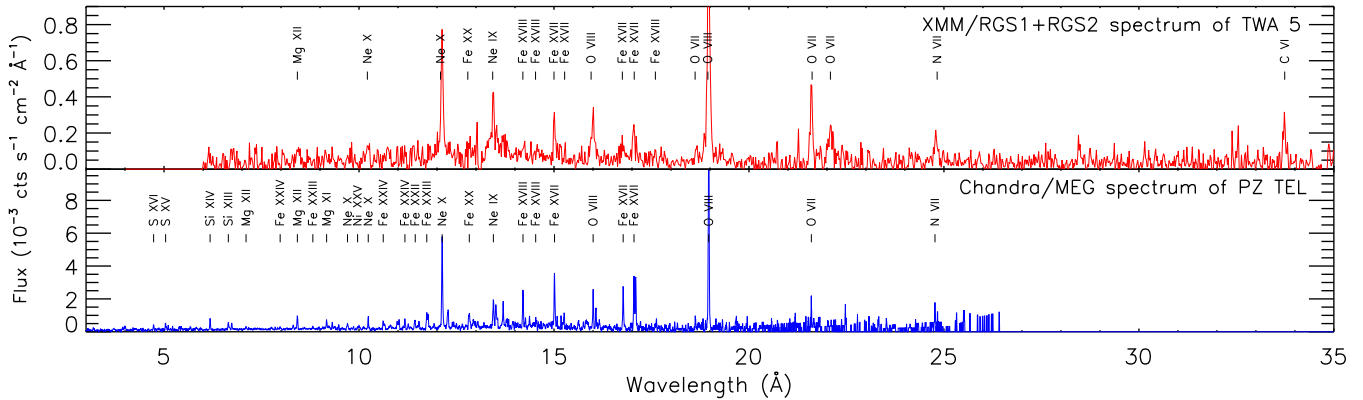}}
\end{center}
\caption{Upper panel: Spectrum of \object{TWA~5} obtained by adding the RGS1 and RGS2 X-ray spectra previously rebinned by a factor 2. Lower panel: Spectrum of \object{PZ~Tel} obtained by smoothing the Chandra/MEG first order spectra. \label{fig2}}
\end{figure*}

\begin{figure}[!b]
\begin{center}
\resizebox{\hsize}{!}{\includegraphics{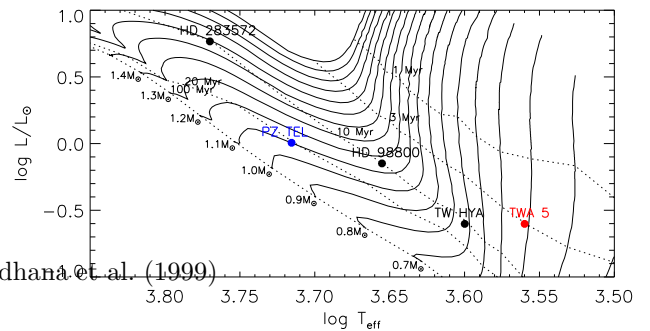}}
\end{center}
\caption{HR diagram with isochrones (dotted lines) and evolutionary tracks (solid lines) from Siess et al. (2000). \label{fig1}}
\end{figure}

\section{Target Information}
\label{target}

The principal characteristics of \object{TWA~5} and \object{PZ~Tel} are reported in Table~\ref{tab1}, together with those of the other comparison PMS stars mentioned above. Note that  \object{TWA~5} is a multiple system: the primary, \object{TWA~5}A, is a spectroscopic binary consisting of similar component, while \object{TWA~5B}, the secondary, is a M8 brown dwarf. \object{HD~98800} is a quadruple system: it consists of two visible components, each of which is a spectroscopic binary. \object{HD~98800} spectral type in Table~\ref{tab1} refers to the two stars which contribute to the X-ray emission. In Figure~\ref{fig1} we show the location of the stars in our sample on the HR~diagram.

\section{Observations}

\object{TWA~5} was observed with {\em XMM-Newton} on 2003 January 9, for 29.7~ks. RGS first order spectra were extracted with SAS~V5.4.1. As shown by \cite*{TsuboiMaeda2003} the X-ray emission of \object{TWA~5} is dominated by the primary component. \object{PZ~Tel} was observed with {\em Chandra}/HETGS on 2003 June 7, for 73.9~ks, and we have used CIAO~V3.0 to extract first order spectra. In Figure~\ref{fig2} both spectra are shown with labels indicating the strongest emission lines. The observed X-ray luminosities, in the range 6-20~\AA, for \object{TWA~5} and \object{PZ~Tel} are $4.4\times10^{29}$ and $2.2\times10^{30}~{\rm erg~s^{-1}}$, respectively.

\section{Results}

We have reconstructed the plasma emission measure distribution ($EMD$) and derived the coronal abundances for both stars applying the MCMC method (\cite{KashyapDrake1998}) based on the measured line fluxes. The results obtained are shown in Figures~\ref{fig3} and \ref{fig4}. The continuum adopted for the line fitting agrees with the continuum predicted by the $EMD$ and the abundance set for each star. The analysis of \object{PZ~Tel} has been performed using the CHIANTI database, while for \object{TWA~5} we used the APED database. Note however that these databases provide compatible results with the analysis approach we have followed, as discussed by \cite*{ScelsiMaggio2004b}.

\subsection{$EMDs$}
\label{emd}

\begin{figure}[ht]
\begin{center}
\resizebox{\hsize}{!}{\includegraphics{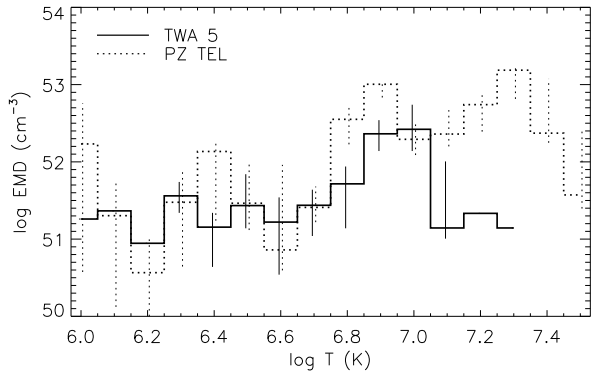}}
\end{center}
\caption{Emission measure distribution of \object{TWA~5} (solid line) and \object{PZ~Tel} (dotted line). \label{fig3}}
\end{figure}

The emission measure distributions that we have derived for both targets (Figure~\ref{fig3}) show characteristics similar to those of other active stars, and in particular a peak at $\log T \sim 6.9\div7.0$ and steep slopes in the temperature range just preceding the peak. Moreover \object{PZ~Tel} has a significant amount of plasma at higher temperatures, with another peak at $\log T=7.3$. The fact that we do not see such hot temperatures in the $EMD$ of \object{TWA~5} may be due to a lower average coronal temperature, or to the different diagnostics offered by {\em XMM-Newton}/RGS in comparison with {\em Chandra}/HETGS. The analysis of {\em XMM-Newton}/EPIC data of \object{TWA~5}, which will appear in a subsequent paper, will help to answer this question.

\subsection{Abundances}
\label{abund}

\begin{figure}[ht]
\begin{center}
\resizebox{\hsize}{!}{\includegraphics{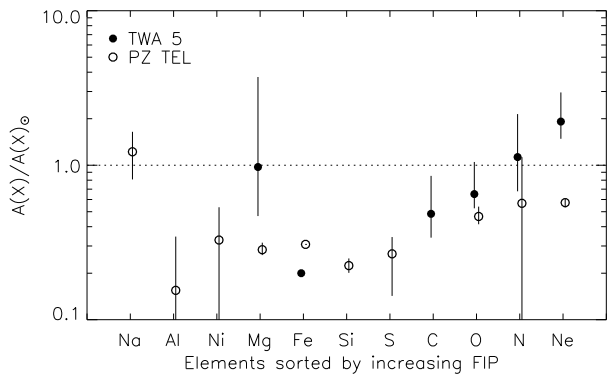}}
\end{center}
\caption{Elemental abundances of \object{TWA~5} (filled symbols) and \object{PZ~Tel} (open symbols) referred to the solar photospheric values (Grevesse \& Sauval 1998). \label{fig4}}
\end{figure}

In Figure~\ref{fig4} we show the derived coronal abundances relative to the solar photospheric values (\cite{GrevesseSauval1998}), this is because photospheric abundances of the relevant stars are not known, and on principle they may differ significantly from solar values. The elements are sorted by increasing first ionization potential (FIP) values. The $EMD$ reconstruction based on line fluxes provides relative abundances of elements for which we have measurable lines. We then obtained abundances relative to H by scaling the metallicity so as to match predicted and observed continuum level in line-free spectral region. For both stars we found a pattern with a minimum abundance at intermediate FIP elements, while the elements with higher FIP show higher values of abundance. This behavior is more evident for \object{TWA~5} whose Ne/Fe abundance ratio is $\sim10$ times higher than the solar one.

\subsection{Densities}
\label{dens}

\begin{table}[!b]
\caption{Electron Densities from He-like triplets.}
\label{tab2}
\begin{center}
\leavevmode
\scriptsize
\begin{tabular}{llccc}
\hline
star    & ion    & $\log T_{\rm max}~{\rm (K)}$ & $f/i$       & $\log N_{\rm e}~({\rm cm^{-3}})$ \\
\hline
\object{TWA~5}   & O~VII  & 6.3                & $3.8\pm2.5$ & $<11$        \\
\object{PZ~Tel}  & Ne~IX  & 6.6                & $3.0\pm1.4$ & $<11.8$      \\
\object{PZ~Tel}  & Mg~XI  & 6.8                & $1.7\pm0.8$ & $12.6\pm0.6$ \\
\object{PZ~Tel}  & Si~XII & 7.0                & $3.5\pm2.1$ & $<13.5$      \\
\hline
\end{tabular}
\end{center}
\end{table}

\begin{figure}[ht]
\begin{center}
\resizebox{\hsize}{!}{\includegraphics{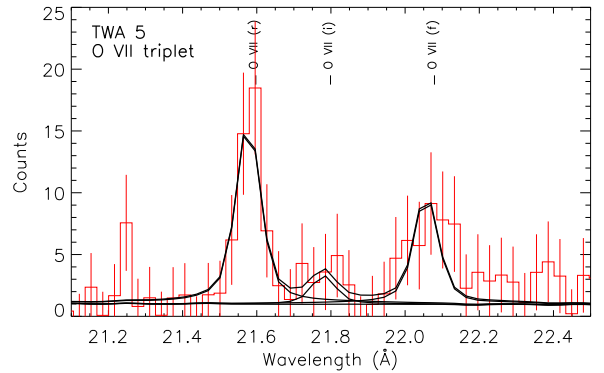}}
\end{center}
\caption{O~VII He-like triplet in the \object{TWA~5} spectrum with best fit curves superimposed. \label{fig5}}
\end{figure}

\begin{figure}[ht]
\begin{center}
\resizebox{\hsize}{!}{\includegraphics{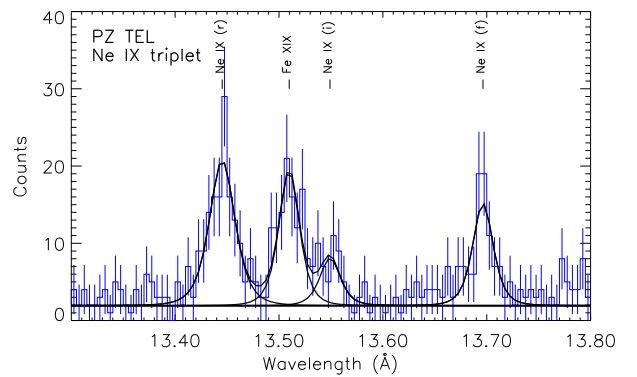}}
\end{center}
\caption{Ne~IX He-like triplet in the \object{PZ~Tel} spectrum with best fit curves superimposed. \label{fig6}}
\end{figure}

We have used the density sensitivity of the intercombination and forbidden lines of He-like ions (\cite{GabrielJordan1969}) to investigate the coronal electron densities $N_{\rm e}$. In the observed X-ray spectrum of \object{TWA~5} we have analyzed only the O~VII triplet, because the Ne~IX triplet is blended with Fe~XIX lines and the RGS spectral resolution is insufficient to resolve these blends. On the other hand in the observed X-ray spectrum of \object{PZ~Tel} we have analyzed the Ne~IX, Mg~XI and Si~XIII triplets, while of the O~VII triplet only the resonance line has been detected. The O~VII He-like triplet for \object{TWA~5} and the Ne~IX triplet for \object{PZ~Tel} are shown in Figures~\ref{fig5} and \ref{fig6}, including observed spectra and best-fit line profiles; the derived $N_{\rm e}$ values are shown in Table~\ref{tab2}. We derive only upper limits for the electron densities from the observed $f/i$ ratio for all triplets except the Mg~XI for \object{PZ~Tel}. It is worth noting that these results are different from those of \object{TW~Hya} where high densities ($N_{\rm e}\sim 10^{13}~{\rm cm^{-3}}$) have been deduced from the analysis of Ne~IX and O~VII triplets (\cite{KastnerHuenemoerder2002,StelzerSchmitt2004}).

\section{Discussion}
\label{disc}

The shape of $EMDs$, the abundance patterns and the electron densities derived from the X-ray spectra of \object{TWA~5} and \object{PZ~Tel} suggest that the X-ray emission from the PMS stars analyzed is very similar to the coronal emission of more evolved active stars. In the following we compare our results with those of \object{TW~Hya}, \object{HD~98800} and \object{HD~283572}.

\begin{figure}[!t]
\begin{center}
\resizebox{\hsize}{!}{\includegraphics{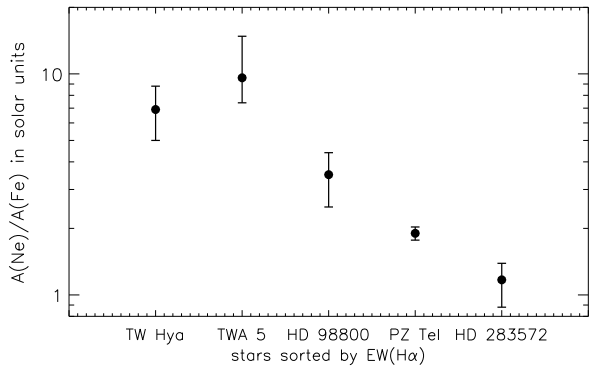}}
\end{center}
\caption{Ne/Fe abundance ratio in solar units (Grevesse \& Sauval 1998) for the selected pre-main sequence stars ordered by decreasing H$\alpha$ emission. We have assumed that the abundance uncertainties of \object{HD~98800} are equal to those of \object{TW~Hya}, because of the similar S/N ratio of the Chandra spectra of these two stars. \label{fig7}}
\end{figure}

\subsection{Ne/Fe abundance ratio}

In Figure~\ref{fig7} we compare the Ne/Fe abundance ratios (referred to the solar value) for these five PMS stars sorted by H$\alpha$ equivalent width. Note however that we obtain a similar pattern if our stars are sorted by spectral type. \object{TWA~5} shows the same ratio as \object{TW~Hya}, while the ratio is significantly lower for the other stars, characterized by no evidence of accretion. This finding suggests that the Ne/Fe abundance ratio may reflect the evolutionary stage of the relevant star, although this interpretation is complicated by the fact that more evolved stars, like the ZAMS K0 dwarf \object{AB~Dor}, show again very high Ne/Fe values ($\sim5\div10$). While Ne/Fe ratios are also thought to be correlated with stellar magnetic activity, another possibility is that the higher Ne/Fe ratio of \object{TW~Hya}, \object{TWA~5} and \object{HD~98800}, which all belong to the TW Hydrae association, reflects the initial abundances of the cloud from which these stars originated, although this would require rather large variations in local ISM abundances.

\subsection{$EMD$ peak temperature}

\begin{figure}[!t]
\begin{center}
\resizebox{\hsize}{!}{\includegraphics{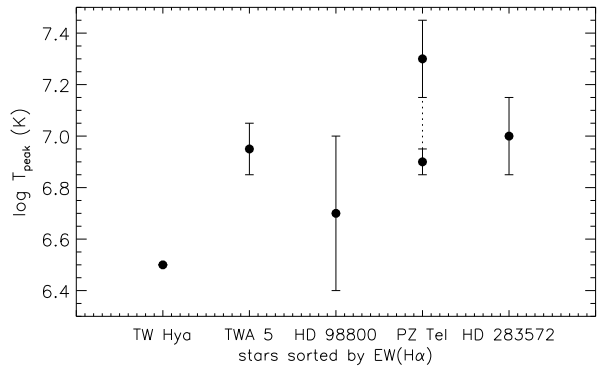}}
\end{center}
\caption{Peak temperature of the $EMD$ for the selected pre-main sequence stars ordered by decreasing H$\alpha$ emission. \label{fig8}}
\end{figure}

In Figure~\ref{fig8} we show the $EMD$ peak temperatures of the same stars. Note that the \object{HD~98800} value is only indicative because no detailed emission measure analysis has been performed on its data. This plot suggests that the temperature of the emitting plasma rises with the ending of the accretion process, although there are other parameter differences between the selected stars.

\section{Conclusions}

In conclusion we have found that Ne/Fe abundance ratio seems to be correlated with evolutionary stages and with spectral type. In particular \object{TWA~5} has the same Ne/Fe abundance ratio as \object{TW~Hya}. On the other hand, none of the sample stars shows the same densities and plasma temperatures as \object{TW~Hya}. We can therefore depict two scenarios: in the first case \object{TW~Hya} may represent CTTSs with ages $\sim10$~Myr, but still characterized by high accretion levels, while the other sample stars represent different evolutionary phases. In the second case \object{TW~Hya}, and/or its environment, may be quite different from the other CTTSs, and therefore it is not straightforward to reconcile its characteristics with those of the other PMS stars.

\begin{acknowledgements}

CA, AM, GP, and SS acknowledge partial support for this work by ASI and MIUR. CA acknowledges ESA for travel support.

\end{acknowledgements}


\begin{thebibliography}{}

\bibitem[\protect\astroncite{Argiroffi et~al.}{2004}]{ArgiroffiDrake2004}
{Argiroffi} C., {Drake} J.~J., {Maggio} A., {Peres} G., {Sciortino} S., 
  {Harnden} F.~R., 2004, ApJ, 609, 925

\bibitem[\protect\astroncite{Gabriel \& Jordan}{1969}]{GabrielJordan1969}
{Gabriel} A.~H., {Jordan} C. 1969, MNRAS, 145, 241

\bibitem[\protect\astroncite{Grevesse \& Sauval}{1998}]{GrevesseSauval1998}
{Grevesse}, N., {Sauval}, A.~J. 1998, Space Science Reviews, 85, 161

\bibitem[\protect\astroncite{Jayawardhana et~al.}{1999}]{JayawardhanaHartmann1999}
{Jayawardhana} R., {Hartmann} L., {Fazio} G., {Fisher} R.~S., {Telesco}
  C.~M., {Pi{\~ n}a} R.~K., 1999, ApJL, 521, L129

\bibitem[\protect\astroncite{Kashyap \& Drake}{1998}]{KashyapDrake1998}
{Kashyap} V., {Drake} J.~J., 1998, ApJ, 503, 450

\bibitem[\protect\astroncite{Kastner et~al.}{2004}]{KastnerHuenemoerder2004}
{Kastner} J.~H., {Huenemoerder} D.~P., {Schulz} N.~S., {Canizares} C.~R.,
  {Li} J., {Weintraub} D.~A., 2004, ApJL, 605, L49

\bibitem[\protect\astroncite{Kastner et~al.}{2002}]{KastnerHuenemoerder2002}
{Kastner} J.~H., {Huenemoerder} D.~P., {Schulz} N.~S., {Canizares} C.~R.,
 {Weintraub} D.~A., 2002, ApJ, 567, 434

\bibitem[\protect\astroncite{Metchev et~al.}{2004}]{MetchevHillenbrand2004}
{Metchev} S.~A., {Hillenbrand} L.~A., {Meyer} M.~R., 2004, ApJ, 600, 435

\bibitem[\protect\astroncite{Mohanty et~al.}{2003}]{MohantyJayawardhana2003}
{Mohanty} S., {Jayawardhana} R., {Barrado y Navascu{\' e}s} D., 2003,
  ApJL, 593, L109

\bibitem[\protect\astroncite{Scelsi et~al.}{2004a}]{ScelsiMaggio2004a}
{Scelsi} L., {Maggio} A., {Peres} G., 2004a, submitted

\bibitem[\protect\astroncite{Scelsi et~al.}{2004b}]{ScelsiMaggio2004b}
{Scelsi} L., {Maggio} A., {Peres} G., {Gondoin} P., 2004b,
  A\&A, 413, 643

\bibitem[\protect\astroncite{Siess et~al.}{2000}]{SiessDufour2000}
{Siess} L., {Dufour} E., {Forestini} M., 2000, A\&A, 358, 593

\bibitem[\protect\astroncite{Soderblom et~al.}{1998}]{SoderblomKing1998}
{Soderblom} D.~R., {King} J.~R., \& {Henry} T.~J., 1998, AJ, 116, 396

\bibitem[\protect\astroncite{Stelzer \& Schmitt}{2004}]{StelzerSchmitt2004}
{Stelzer} B., {Schmitt} J.~H.~M.~M., 2004, A\&A, 418, 687

\bibitem[\protect\astroncite{Tsuboi et~al.}{2003}]{TsuboiMaeda2003}
{Tsuboi} Y., {Maeda} Y., {Feigelson} E.~D., {Garmire} G.~P., {Chartas} G.,
  {Mori} K., {Pravdo} S.~H., 2003, ApJL, 587, L51

\bibitem[\protect\astroncite{Weinberger et~al.}{2004}]{WeinbergeBecklinr2004}
{Weinberger} A.~J., {Becklin} E.~E., {Zuckerman} B., {Song} I., 2004, AJ,
  127, 2246

\end{thebibliography}
\end{document}